\documentclass[final,5p,times,twocolumn]{elsarticle}

\usepackage{amssymb}
\usepackage{amsmath}

\usepackage{booktabs}
\usepackage{xcolor}
\usepackage{adjustbox}
\usepackage{caption}

\biboptions{sort&compress}

\journal{Photoacoustics}
\begin{document}

\begin{frontmatter}





\title{Application of a Virtual Imaging Framework for Investigating a Deep Learning-Based Reconstruction Method for 3D Quantitative Photoacoustic Computed Tomography}



\author[label1]{Refik Mert Cam}
\author[label2]{Seonyeong Park}
\author[label3,label4]{Umberto Villa}
\author[label1,label2]{Mark A. Anastasio}

\affiliation[label1]{organization={Department of Electrical \& Computer Engineering, University of Illinois Urbana-Champaign},
            postcode={61801}, 
            state={IL},
            country={USA}}

\affiliation[label2]{organization={Department of Bioengineering, University of Illinois Urbana-Champaign},
            postcode={61801}, 
            state={IL},
            country={USA}}

\affiliation[label3]{organization={Oden Institute for Computational Engineering and Sciences, The University of Texas at Austin},
            postcode={78712}, 
            state={TX},
            country={USA}}
            
\affiliation[label4]{organization={Department of Biomedical Engineering, The University of Texas at Austin},
            postcode={78712}, 
            state={TX},
            country={USA}}
\begin{abstract}
Quantitative photoacoustic computed tomography (qPACT) is a promising imaging modality for estimating physiological parameters such as blood oxygen saturation. However, developing robust qPACT reconstruction methods remains challenging due to computational demands, modeling difficulties, and experimental uncertainties. Learning-based methods have been proposed to address these issues but remain largely unvalidated. Virtual imaging (VI) studies are essential for validating such methods early in development, before proceeding to less-controlled phantom or in vivo studies. Effective VI studies must employ ensembles of stochastically generated numerical phantoms that accurately reflect relevant anatomy and physiology. Yet, most prior VI studies for qPACT relied on overly simplified phantoms. In this work, a realistic VI testbed is employed for the first time to assess a representative 3D learning-based qPACT reconstruction method for breast imaging. The method is evaluated across subject variability and physical factors such as measurement noise and acoustic aberrations, offering insights into its strengths and limitations.

\end{abstract}



\begin{keyword}
Quantitative photoacoustic computed tomography, numerical breast phantoms, breast imaging, virtual imaging studies




\end{keyword}

\end{frontmatter}


\section{Introduction}
Photoacoustic computed tomography (PACT) is an emerging non-invasive modality that offers high spatial resolution and optical contrast~\cite{wang2003noninvasive, wang2012photoacoustic, wang2015photoacoustic, wang2017photoacoustic}. 
PACT is employed for structural and functional imaging of biological tissues across preclinical and clinical contexts~\cite{wang2003noninvasive, wang2012photoacoustic, wang2015photoacoustic, wang2017photoacoustic, lozenski2024proxnf, cam2024spatiotemporal}.
It is a hybrid imaging technique that combines optical excitation and ultrasonic detection, leveraging the photoacoustic effect, where absorbed optical energy causes rapid thermoelastic expansion, resulting in the generation of acoustic waves \cite{wang2012photoacoustic, wang2015photoacoustic}. These acoustic waves then propagate through tissue and are detected by an array of ultrasonic transducers positioned around the imaging target. The recorded signals are subsequently employed for image reconstruction, enabling visualization of the spatial distribution of absorbed optical energy. By using PACT measurements acquired at multiple excitation wavelengths, it is, in principle, possible to estimate absolute or relative physiological quantities (e.g., blood oxygen saturation) and molecular quantities (e.g., concentrations of chromophores) within biological tissue~\cite{rosenthal2009quantitative, ntziachristos2010molecular, tzoumas2016eigenspectra, bal2011multi, bal2012multi}. This technique is referred to as quantitative PACT (qPACT)~\cite{cox2009challenges, bal2011multi, bal2012multi, rosenthal2009quantitative}. 

The qPACT inverse problem is nonlinear and inherently ill-posed because of the coupled physics of light transport and photoacoustically induced pressure generation. Even under ideal, noise-free conditions, 
different combinations of optical absorption, optical scattering, and the Gr\"uneisen parameter can yield indistinguishable measurement data, leading to non-uniqueness and instability in the inversion~\cite{bal2011multi,bal2012multi,cox2009challenges}. Beyond this fundamental limitation, in practical cases, the difficulty of the qPACT inverse problem is further exacerbated due to multiple factors such as imperfect system characterization, model mismatch in the optical and acoustic forward models (e.g., uncertainty in heterogeneous optical and acoustic properties), and limited angular/aperture coverage~\cite{bal2011multi,bal2012multi,cox2009challenges}. Physics-based reconstruction methods with advanced regularization and learning-based methods have been proposed to address these challenges. However, the development of accurate and robust image reconstruction methods that are suitable for deployment in practice remains an active research topic
~\cite{javaherian2019direct, mamonov2012quantitative, bal2011multi, bal2012multi, bench2020toward, luke2019net, cai2018end, chen2020deep, grohl2019estimation, yang2019eda, yang2019quantitative, durairaj2020unsupervised, liang2025self, cam2024investigation}.

The development of rigorous 
evaluation frameworks is essential for advancing qPACT reconstruction methods. \emph{In vivo} data generally lack ground truth of to-be-estimated quantities, which makes them unsuitable for quantitative evaluation. Physical phantoms offer controlled imaging conditions but are often overly simplistic and typically lack anatomical and physiological realism~\cite{else2024effects, durairaj2020unsupervised, rosenthal2009quantitative}. Moreover, fabricating large numbers of physical phantoms that realistically represent 
clinically relevant variability, such as acoustic heterogeneity, anatomical realism, and physiological complexity, can be prohibitively costly and impractical~\cite{kiarashi2015development, keenan2016design}.

Virtual imaging (VI) studies (i.e., computer-simulation studies that pair realistic numerical phantoms with high-fidelity forward models of data acquisition) offer an alternative principled route to such quantitative evaluations~\cite{park2023numerical, park2023stochastic, cam2024investigation, chen2025benchmarking}. In the context of qPACT, VI enables independent control of optical and acoustic parameters, acquisition geometry, noise, and reconstruction assumptions while preserving access to reference optical/functional maps. To be effective, VI studies require ensembles of numerical phantoms that capture clinically relevant anatomical and physiological variability and that support stochastic assignment of tissue-specific optical and acoustic properties. When designed in this way, VI studies can quantify performance across a cohort of virtual subjects, reveal failure modes, and guide algorithm design and translation~\cite{park2023numerical, park2023stochastic, cam2024investigation, chen2025benchmarking}.

This work employs a realistic VI framework based on ensembles of anatomically and physiologically realistic three-dimensional (3D) numerical breast phantoms (NBPs)~\cite{park2023stochastic, park2023numerical} to enable the systematic and quantitative assessment of a qPACT reconstruction method. To our knowledge, this is the first time that a realistic VI testbed has been employed for this purpose. 
Specifically, a 3D learning-based qPACT method for breast imaging is systematically evaluated with consideration of an ensemble of to-be-imaged subjects and physical factors that include measurement noise and acoustic aberration in the measurement data. Two VI studies, each based on distinct modeling assumptions, are designed to assess robustness and generalization across a range of object-level variations. These include spatial heterogeneity in acoustic properties (sound speed, density, and attenuation), anatomical differences in breast size and tissue composition, as well as optical variations in skin tone.
The resulting analyses reveal strengths and limitations of the considered learned qPACT method and, more importantly, demonstrate the value of realistic VI studies for accelerating the development and facilitating the validation of effective qPACT image reconstruction methods.

The remainder of this paper is organized as follows. Section~\ref{background} summarizes the imaging physics of qPACT and reviews reconstruction approaches. 
Section~\ref{method} presents the VI framework and describes the evaluation of a representative deep learning (DL)-based qPACT method using NBPs, realistic imaging conditions, and clinically motivated study designs. Section~\ref{results} reports the results of VI studies. Finally, Section~\ref{discussion_and_conclusion} presents a combined discussion and conclusion, including limitations and directions for future work.

\section{Background}
\label{background}
\subsection{Imaging physics of quantitative PACT}
\label{sec:fwd_qpact}
In PACT, a short laser pulse illuminates the object-to-be-imaged (typically biological tissue).
Absorption of optical energy by various chromophores (light-absorbing molecules) within the object induces a localized increase in acoustic pressure through the photoacoustic effect~\cite{wang2017photoacoustic, wang2015photoacoustic, wang2012photoacoustic}. 
Mathematically, the induced initial pressure distribution $p_0(\mathbf{r}, \lambda)$ at position $\mathbf{r}\in\mathbb{R}^3$ and excitation wavelength $\lambda$ is expressed as \cite{cox2009challenges, beard2011biomedical, cox2006two}:
\begin{equation}
p_0(\mathbf{r}, \lambda) = \Gamma\, A(\mathbf{r}, \lambda) = \Gamma\, \mu_a(\mathbf{r}, \lambda)\, \Phi(\mathbf{r}, \lambda; \mu_{a}, \mu_{s}, g, n).
\label{eq:initial_pressure}
\end{equation}
Here, $A(\mathbf{r}, \lambda)$, $\mu_a(\mathbf{r}, \lambda)$, and $\Phi(\mathbf{r}, \lambda; \mu_{a}, \mu_{s}, g, n)$ are the wavelength-dependent absorbed optical energy, optical absorption coefficient, and optical fluence, respectively, and $\Gamma$ is the Gr\"uneisen parameter that describes the conversion efficiency from absorbed optical energy to acoustic pressure. The optical fluence is dependent on the tissue’s optical properties, specifically the absorption coefficient $\mu_a(\mathbf{r}, \lambda)$, the scattering coefficient $\mu_s(\mathbf{r}, \lambda)$, the scattering anisotropy factor $g(\mathbf{r}, \lambda)$, and the refractive index $n(\mathbf{r}, \lambda)$.

The optical absorption coefficient is determined by the concentrations of various chromophores present in the tissue \cite{cox2009challenges, jacques2013, park2023numerical, park2023stochastic}:
\begin{equation}
\mu_a(\mathbf{r}, \lambda) = \sum_{k\in\mathcal{K}} c_k(\mathbf{r})\, \varepsilon_k(\lambda),
\label{eq:absorption_coefficient}
\end{equation}
where $c_k(\mathbf{r})$ denotes the molar concentration of chromophore $k$ at position $\mathbf{r}$, and $\varepsilon_k(\lambda)$ is the corresponding molar extinction coefficient at wavelength $\lambda$. The set $\mathcal{K}$ denotes the chromophores in the object. Key chromophores in biological tissues within the optical wavelengths relevant to PACT include oxyhemoglobin ($\text{HbO}_2$), deoxyhemoglobin (Hb), melanin, lipids and water~\cite{jacques2013}.

Once the initial pressure is induced, it serves as the source of acoustic wave propagation. The resulting acoustic wavefield propagates through the medium and is recorded by ultrasonic transducers~\cite{wang2017photoacoustic, wang2015photoacoustic, wang2012photoacoustic}. The recorded data can then be used to reconstruct the initial pressure distribution and, in the context of qPACT, to estimate spatial distributions of tissue optical properties and/or molecular constituents~\cite{bench2020toward, luke2019net, cai2018end, chen2020deep, grohl2019estimation, yang2019eda, yang2019quantitative, durairaj2020unsupervised, liang2025self, cam2024investigation}. This typically involves acquiring measurements under multiple different optical excitation conditions, most commonly by varying the illumination wavelength~\cite{cox2009challenges, bal2011multi, bal2012multi, wang2023optical}. The goal of qPACT may include recovering absolute or relative values of optical absorption coefficients, scattering properties, or concentrations of specific chromophores~\cite{cox2006two, cox2009challenges, bal2011multi, bal2012multi, wang2023optical, cam2024investigation}.

\subsection{Inversion methods for qPACT}
Linear spectral unmixing, while not an accurate method, is nevertheless commonly employed for quantitative estimation from multispectral PACT measurements~\cite{tzoumas2017spectral, park2022normalization, wang2023optical}. This method simplifies the nonlinear inverse problem to a linear one, neglecting wavelength-dependent optical fluence variations caused by differential absorption and scattering during light propagation in the object, 
known as spectral coloring effects~\cite{cox2009challenges, beard2011biomedical, bench2020toward, tzoumas2016eigenspectra, ntziachristos2010molecular}. These effects become increasingly significant at greater depths, where cumulative absorption and scattering degrade accuracy
~\cite{cox2009challenges, cox2009estimating, tzoumas2016eigenspectra, ntziachristos2010molecular}.
To address this, physics-model-based inversion techniques have been developed~\cite{javaherian2019direct, mamonov2012quantitative, cox2009estimating, bal2011multi, bal2012multi, tzoumas2016eigenspectra}. These methods incorporate detailed models of light propagation in biological tissues and employ carefully devised regularization schemes to address the ill-posed nature of the problem~\cite{javaherian2019direct, mamonov2012quantitative, cox2009estimating, bal2011multi, bal2012multi}. Despite their potential, physics-based qPACT methods face several challenges that limit their clinical applicability, including high computational demands, sensitivity to modeling errors, and the difficulties in designing 
robust regularization strategies to handle parameter uncertainty and incomplete or noisy data~\cite{tarvainen2024quantitative, ren2020characterizing, fonseca2016sensitivity, wang2023optical}.

DL-based approaches offer an alternative solution by leveraging data-driven models to approximate the mapping from photoacoustic measurements to tissue optical and/or functional properties~\cite{bench2020toward, luke2019net, cai2018end, chen2020deep, grohl2019estimation, yang2019eda, yang2019quantitative, durairaj2020unsupervised, liang2025self, cam2024investigation, li2022deep, grohl2019estimation}. Among these, convolutional neural networks (CNNs) represent one of the most widely used architectures and have been employed in qPACT methods to learn this mapping~\cite{bench2020toward, luke2019net, cai2018end, chen2020deep, yang2019eda, liang2025self, cam2024investigation}. However, most existing studies have been conducted using only 
simplified numerical and/or physical phantoms, both of which lack anatomical and physiological realism~\cite{bench2020toward, luke2019net, cai2018end, chen2020deep, yang2019eda, liang2025self, cam2024investigation}. Additionally, the majority of these works focus on two-dimensional (2D) imaging scenarios~\cite{luke2019net, cai2018end, chen2020deep, yang2019eda, liang2025self, else2024effects}.
Even the limited number of studies that explored 3D tomographic imaging using VI studies 
employed simplified numerical phantoms that do not accurately capture realistic heterogeneity in tissue properties~\cite{bench2020toward, cam2024investigation}. As a result, the performance of DL-based qPACT methods under clinically relevant scenarios remains insufficiently evaluated~\cite{else2024effects, cam2024investigation}.
In particular, robustness to epistemic uncertainty, which stems from limited knowledge of the to-be-imaged object, including generally inaccessible, spatially heterogeneous acoustic properties such as speed-of-sound (SOS), is a critical yet underexplored factor that can significantly impact estimation accuracy.
This challenge is compounded by the fact that \emph{in vivo} experimental imaging data generally lack reference values for optical and functional parameters, making rigorous validation difficult. Therefore, there is a critical need for VI studies that reflect realistic and clinically relevant variability, for systematic and quantitative evaluation of DL-based qPACT reconstruction methods.

\section{Evaluation of a DL-based qPACT method using a virtual imaging framework} 
\label{method}
A representative DL–based qPACT reconstruction method is evaluated using 
a realistic VI framework for controlled, quantitative assessment across cohorts of virtual subjects. The VI setup, comprising the multispectral photoacoustic data simulation pipeline, together with an ensemble of 3D NBPs, is described in Section~\ref{virtual_imaging_framework}. The DL method, including network architecture, loss functions, and the training protocol with data augmentation, is specified in Section~\ref{DL-based_qPACT_method}. Two study designs are introduced to probe robustness and generalization under clinically relevant variability in Section~\ref{virtual_imaging_study_designs}.

\subsection{Virtual imaging framework}
\label{virtual_imaging_framework}
The VI framework comprises two essential components: (i) an ensemble of anatomically and physiologically realistic 3D NBPs that provide known optical, acoustic, and functional maps with population variability for quantitative evaluation, and (ii) a simulation pipeline configured with a VI system emulating a hemispherical breast PACT imager with multispectral illumination.

\subsubsection{Stochastic numerical breast phantoms}
\label{nbps}
\begin{figure*}[ht]
\centering
\includegraphics[width=0.99\linewidth]{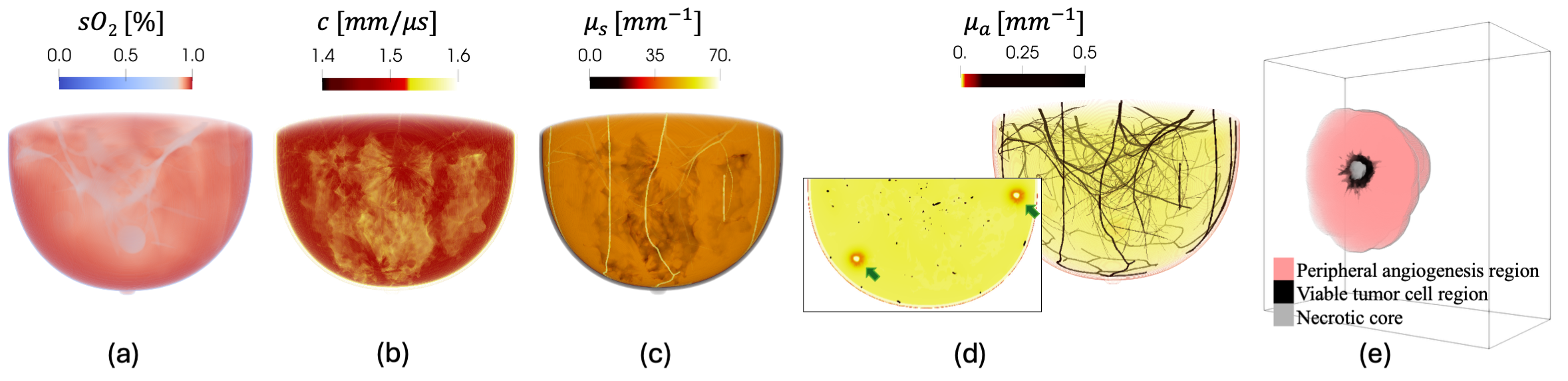}
\caption{Distributions of functional, acoustic, and optical properties of a representative type B NBP with an embedded malignant tumor: 
(a) blood oxygen saturation $sO_2$, (b) speed of sound $c$, (c) optical scattering coefficient $\mu_s$ at a wavelength of 757~nm, (d) optical absorption coefficient $\mu_a$ at 757~nm, and (e) 3D malignant tumor model. For visualization purposes, the tumor is shown as a split volume in (e). The inset in (d) displays a cross-section with arrows indicating the tumor locations. Volumetric renderings were generated using ParaView~\cite{ParaView}, and color maps were manually adjusted to enhance visual clarity. \textbf{These anatomically realistic numerical phantoms provide a versatile and clinically meaningful platform for developing and evaluating qPACT techniques under realistic physiological and anatomical variability.}}
\label{phantom_illustration}
\end{figure*}

3D NBPs were generated using a stochastic framework ~\cite{park2023stochastic,park2023numerical} that produces anatomically and physiologically realistic cohorts spanning breast size and shape, tissue composition across BI-RADS density categories (A–D)~\cite{american2013acr}, realistic vasculature, and skin tone (Fitzpatrick 1-6). Unlike simplified models, these NBPs assign tissue-specific, literature-informed heterogeneous optical (e.g., wavelength-dependent absorption and scattering), acoustic (e.g., speed of sound, density, and attenuation), and functional (e.g., blood oxygen saturation) property maps. The framework also permits insertion of anatomically realistic tumors at physiologically plausible locations; malignant tumors are represented with 
a distinct viable tumor cell region exhibiting spiculated morphology, along with a necrotic core and a peripheral angiogenesis region~\cite{park2023stochastic}.
Representative property distributions of an NBP and the tumor model are shown in Fig.~\ref{phantom_illustration}. These phantoms provide the controlled heterogeneity and reference values required for cohort-level, reproducible assessments within the VI studies.

\subsubsection{Virtual imaging system and data simulation}
\label{virtual_imaging_system_configuration}
\begin{figure}[h!]
\centering
\includegraphics[width=0.99\linewidth]{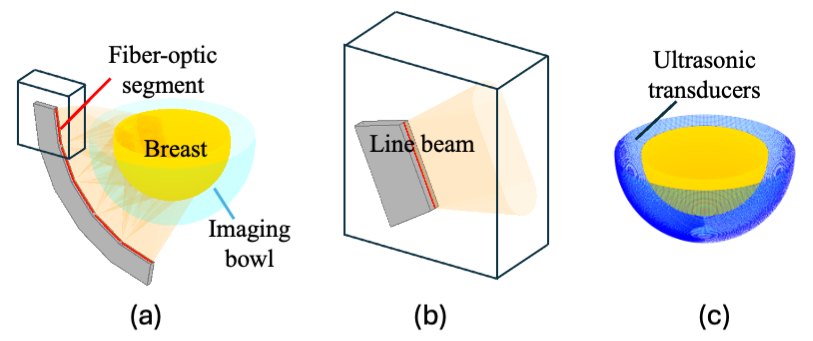}
\caption{Virtual imaging system configuration. (a) Arc-shaped light delivery subsystem composed of linear fiber-optic segments; (b) schematic of a custom line beam with conical angular emission from 
a single fiber-optic segment; (c) all effective ultrasonic transducer positions from 
the rotating arc‐shaped array around the breast across 480 
tomographic view steps
~\cite{louisa,park2023stochastic}.}
\label{vi_system_visualization}
\end{figure}

A VI system was configured to closely emulate an existing breast PACT imaging system, as illustrated in Fig.~\ref{vi_system_visualization}~\cite{louisa, park2023stochastic}. The optical delivery subsystem comprised 20 arc-shaped illuminators (each spanning 80$^\circ$) uniformly arranged on a 145\,mm-radius hemispherical shell around 
the $z$-axis. 
Each illuminator contained five linear fiber-optic segments, producing a total of 100 custom line beams with a conical angular distribution characterized by a half-angle of 12.5$^\circ$; 
further details can be found in~\cite{park2023stochastic,chen2025benchmarking}. The acoustic detection subsystem was equipped with 
108 idealized point-like transducers uniformly distributed on a rotating 85\,mm-radius, 80$^\circ$ arc-array. 
In this configuration, each transducer recorded 3720 temporal samples at a 20~MHz sampling frequency across 480 evenly distributed tomographic views; 
see~\cite{park2023stochastic} for further details.

Synthetic measurement data were generated in two stages. First, the induced initial pressure distribution in Eq. \eqref{eq:initial_pressure} was simulated 
using the GPU-accelerated Monte Carlo eXtreme (MCX, v1.9.0)~\cite{mcx1, mcx2} software to model photon transport at three wavelengths (757, 800, and 850~nm). The Gr\"{u}neisen parameter $\Gamma$ was set to 1, as often assumed for soft tissue~\cite{park2023stochastic}. 
Second, the subsequent propagation and detection of pressure waves were simulated using the k-Wave GPU toolbox~\cite{k-wave}. Transducer positions were approximated by assigning them to the nearest voxel centers on the acoustic simulation grid, discretized with voxel size of 0.25~mm.

\subsection{DL-based qPACT method}
\label{DL-based_qPACT_method}
\begin{figure*}[ht]
\centering
\includegraphics[width=0.7\linewidth]{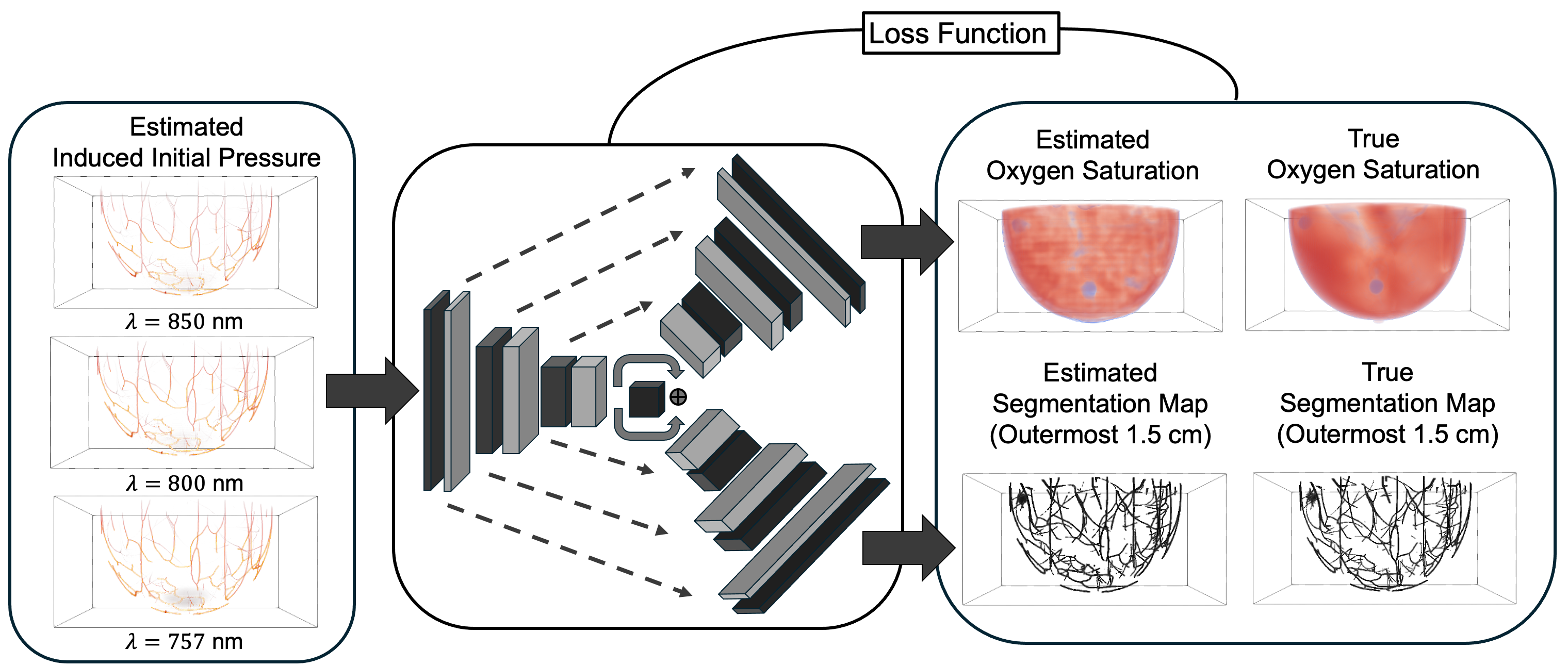}
\caption{Overview of the dual-task DL network for simultaneous $\text{sO}_2$ estimation and anatomical segmentation in 3D photoacoustic tomographic images. Three reconstructed initial pressure distributions at illumination wavelengths of 757, 800, and 850~nm serve as inputs to a shared encoder. Two separate decoders then generate (i) a whole-breast $\text{sO}_2$ map and (ii) a binary segmentation map restricted to the outermost 1.5~cm shell from the breast surface, where veins, arteries, and tumors (if present) regions are labeled as 1 and all other voxels as 0. 
A combined loss function compares the predicted outputs with the corresponding ground truth maps 
($\text{sO}_2$ maps and segmentation masks).}
\label{proposed_method_schematic}
\end{figure*}

A representative DL-based qPACT method was implemented to simultaneously estimate blood oxygen saturation ($\text{sO}_2$) and segment clinically relevant target anatomical structures, specifically vessels and tumor regions (viable tumor cells), from full-scale 3D breast PACT images.
The framework takes as input reconstructed initial pressure estimates at three illumination wavelengths (757, 800, and 850~nm). It estimates both an $\text{sO}_2$ map and a binary segmentation mask in which arteries, veins, and viable tumor cells (if present) are labeled as 1, and all other voxels as 0.
Segmentation is limited to a 1.5\,cm-thick shell defined by depth from the breast surface, because optical attenuation causes exponential decay of photoacoustic signal intensity with depth, limiting recoverable signal information in deeper regions~\cite{park2022normalization}. 
While estimation and segmentation beyond this depth may be feasible, the 1.5 cm threshold represents an empirical design choice that may be revisited in future studies. 
The method leverages multi-task learning to improve accuracy by exploiting correlations between the $\text{sO}_2$ map and the underlying anatomical structures.
An overview of the dual-task network is illustrated in Fig.~\ref{proposed_method_schematic}.

\subsubsection{Network architecture and loss functions}
The architecture adopts a residual encoder–decoder design with a single residual encoder and two task-specific decoders. The encoder 
extracts features from input 3D PACT images. It consists of five levels, each comprising a single residual block that includes two sequential 3$\times$3$\times$3 convolutional layers with leaky ReLU activations.
At each level, feature map dimension is reduced via 3D max pooling (2$\times$2$\times$2 kernel, stride 2), enabling hierarchical multi-scale feature extraction.
Shortcut connections are realized through 1$\times$1$\times$1 convolutional layers that facilitate residual learning and stabilize gradient propagation.

At the network’s bottleneck, the encoded feature representations are refined via an integrated attention module that combine both spatial and channel attention mechanisms~\cite{mou2021cs2}. Following attention-guided feature enhancement, the network bifurcates into two decoder streams: one dedicated to the segmentation task and the other to the regression ($\text{sO}_2$ estimation) task. Both decoders utilize deconvolutional layers (2$\times$2$\times$2 kernel, stride of 2) for upsampling, interspersed with residual decoding blocks that mirror the encoder’s use of two 3$\times$3$\times$3 convolutional layers and leaky ReLU ($\alpha=0.1$) activations. Shortcut connections are employed at each scale by concatenating encoder outputs with decoder inputs, preserving high-resolution details. The final layer of each decoder applies a 1$\times$1$\times$1 convolution followed by a sigmoid activation.

The network is trained using a composite loss function $\mathcal{L}_\text{total}$ that integrates a regression term for $\text{sO}_2$ estimation and a segmentation term: 
\begin{equation}
\label{eq:total_loss}
\mathcal{L}_\text{total} \;=\; \mathcal{L}_\text{reg} \;+\; \eta \mathcal{L}_\text{seg},
\end{equation}
where \(\mathcal{L}_\text{reg}\) denotes the weighted mean squared error for $\text{sO}_2$ estimation, and \(\mathcal{L}_\text{seg}\) is a combination of voxel-weighted binary cross-entropy and Dice loss for segmentation. The scalar $\eta$ is a tunable hyperparameter that balances 
the regression and segmentation terms. 
Further details of the loss functions are provided in~\ref{appendix-loss-regression} and ~\ref{appendix-loss-segmentation}.

\subsubsection{Training and data augmentation}
The training and validation datasets consisted of NBP pairs generated exclusively with Fitzpatrick skin tone 1. Each pair contained one NBP representing a healthy breast and the corresponding NBP with an inserted tumor, differing in tumor presence while sharing identical breast anatomy. 
This design isolates the effect of the tumor without introducing other anatomical variability. 
The training set included 320 such pairs and was structured to reflect a clinically representative distribution of BI-RADS breast density categories: 10\% each for types A and D, and 40\% each for types B and C~\cite{american2013acr}. The validation set comprised 40 pairs and maintained the same distribution to ensure consistency.

Training was performed using the ADAM 
optimizer~\cite{kingma2017adam} with a step size of \(10^{-5}\) and 
was conducted on two NVIDIA A100 GPUs, each with 80 GB of memory. To reduce training time, data parallelization was implemented, and the batch size was set to 2 due to memory constraints and the complexity of the model. A curriculum-based~\cite{wang2021survey, weinshall2020theory} weighting schedule for the composite loss was employed (\ref{appendix-loss-curriculum}), and training proceeded for a total of 600 epochs.

During training, data augmentation was applied at each epoch. Specifically, NBPs were 
rotated by a randomly chosen integer multiple of \(18^\circ\), matching the angular spacing of the 20-view illumination geometry. This approach exploited the inherent symmetry of the illumination setup~\cite{louisa} and 
ensured that the network was exposed to a diverse set of training samples generated from different orientations, thereby reducing overfitting and enhancing robustness to variations in spatial arrangement of the imaging target.

\subsection{Virtual imaging study designs and evaluation}
\label{virtual_imaging_study_designs}
This section describes the design of two VI studies and the evaluation framework used to assess the DL-based qPACT method. The studies were formulated to examine robustness under varying levels of complexity by introducing modeling discrepancies during the reconstruction of induced initial pressure estimates.
Baseline comparison methods and quantitative evaluation metrics are also presented.

\subsubsection{Study definitions}
Two VI studies were conducted to evaluate the representative DL-based qPACT method described in Section~\ref{DL-based_qPACT_method}.

\textbf{Study 1} represents an idealized scenario where reconstruction artifacts are absent and noise is the only source of image degradation. Instead of performing acoustic reconstruction to generate the input to the DL-based qPACT method, the ground truth induced initial pressure distributions were directly corrupted with colored noise. Specifically, independently and identically distributed (iid) zero-mean Gaussian measurement noise was mapped into the image domain using the time-reversal method~\cite{fink1992time, fink2001acoustic, treeby2010photoacoustic}, assuming a constant SOS of water. This process resulted in colored noise. The standard deviation of the noise distribution was set to 1\% of the ensemble mean of the maximum acoustic signal strength across all three wavelengths (757, 800, and 850~nm), as determined from the simulated acoustic pressure measurements generated for Study 2. 

\textbf{Study 2} represents a more realistic and challenging scenario. The acoustic measurement data were simulated by using NBPs (see Section~\ref{nbps}) that incorporate heterogeneous SOS, acoustic density, and attenuation. The acoustic forward simulation employed grid discretization with 0.25 mm voxels. The resulting simulated pressure data were corrupted with additive iid Gaussian noise, with zero mean and a standard deviation equal to 1\% of the ensemble mean of the maximum acoustic signal strength across all three wavelengths (757, 800, and 850~nm). Following the simulation, time-reversal reconstructions were performed to generate the input to the DL-based qPACT method, which assumed a constant SOS, uniform acoustic density, and the absence of acoustic attenuation. The SOS of the water, the acoustic coupling medium, was assumed. 
A computational grid discretized with a voxel size of 0.3 mm was employed for time reversal reconstruction, introducing grid mismatch. During reconstruction, transducer positions defined on the 0.25~mm forward simulation grid were approximated by the closest voxels on the coarser 0.3~mm grid. When multiple positions mapped to the same location, only one was retained. 
This scenario reflects the complexities encountered in practical imaging environments.

The progression from \textbf{Study 1} to \textbf{Study 2} represents a systematic exploration of the DL-based qPACT method's 
performance under increasingly realistic and adverse conditions, thereby establishing a framework for evaluating the robustness of the reconstruction method.

\subsubsection{Baseline comparison methods}
Two baseline methods were employed to benchmark the DL-based qPACT method: linear spectral unmixing and fluence-compensated linear spectral unmixing~\cite{tzoumas2017spectral, park2022normalization, wang2023optical}. These methods serve as reference standards for evaluating accuracy and robustness in estimating blood oxygen saturation.

The first baseline, linear spectral unmixing, assumes wavelength-invariant optical fluence. The second baseline, fluence-compensated linear spectral unmixing, seeks to reduce errors from this assumption by incorporating estimated optical fluence maps for each wavelength.
This approach assumes prior knowledge of the breast volume segmentation and uniform optical properties (absorption, scattering, anisotropy, and refractive index) within the breast region. 
The property values were computed as ensemble averages from the training dataset, while the water region was 
assigned the corresponding optical properties of water. Fluence maps were generated with MCX simulations and applied to rescale the initial pressure estimates, thus compensating for wavelength-dependent fluence variations before spectral unmixing. Although this method does not fully eliminate errors, it improves accuracy by partially accounting for spatial and spectral variations in light propagation, making it a stronger baseline than standard linear spectral unmixing.

\subsubsection{Evaluation strategy}
A comprehensive evaluation was performed using both qualitative and quantitative analyses on ensembles of NBPs. Two categories of test data were considered: an in-distribution (ID) test set, whose characteristics match the training data and which was used to assess accuracy of the learned model, and out-of-distribution (OOD) test sets, whose characteristics differ from the training data and which was used to evaluate generalizability. The ID test set consisted of 64 NBP pairs, each comprising a breast without a tumor and the corresponding breast with tumors. 
In these test sets, BI-RADS breast density types A-D were evenly distributed, in contrast to the 1:4:4:1 ratio used in training.
The term ``in-distribution" denotes that the test set was generated using the same anatomical parameterization and within the same ranges of optical and acoustic tissue properties as the training set.
A balanced distribution 
of breast density types in the ID test set was intentionally adopted to prevent performance metrics from being skewed by overrepresented categories.
The OOD test sets each consisted of 64 NBPs with Fitzpatrick skin tones 3 (OOD-I) and 5 (OOD-II). They shared identical breast anatomy with the ID set but included only tumor-bearing cases, differing solely in skin pigmentation. These sets were designed to evaluate the robustness of the DL-based qPACT method against real-world variability in skin pigmentation.

Separate evaluation within tumor and vessel regions is critical, as the photoacoustic signal originating from tumors is significantly weaker than that from vascular structures. This disparity necessitates tailored assessment strategies to accurately characterize model performance across these regions. To mitigate class imbalance during training and simplify the segmentation task, the model was designed to produce a unified binary mask encompassing both tumors and vessels. 
Because these structures differ in their morphology, 
effective post hoc separation was feasible. A dedicated post-processing pipeline was implemented to achieve this differentiation. A multiscale Frangi vesselness filter was applied to the segmentation output to enhance vascular features~\cite{frangi1998multiscale}, and the resulting vesselness map was thresholded to generate a binary vessel mask. Connected component analysis was then used to identify contiguous vascular regions, with components classified as vessels if more than 50\% of their voxels were labeled as vessel in the thresholded map. To disjoin adjacent 
tumors and vessels, morphological operations consisting of erosion followed by dilation were applied. 
Minor manual refinements were subsequently performed to correct vessel components that were erroneously labeled as tumors upon visual inspection.

To comprehensively evaluate model performance, targeted assessments were conducted for tumor detection, vascular segmentation, and regional $\text{sO}_2$ estimation. Considering the potential diagnostic application of PACT~\cite{lin2022emerging, lin2021photoacoustic}, the evaluation emphasized tumor detection rather than segmentation accuracy over tumor regions.
Tumor detection was determined by comparing the predicted segmentation to the ground-truth binary tumor mask; 
a tumor was considered detected if the overlap exceeded 500 voxels, and undetected otherwise. The ground-truth tumor region comprised 3,808 voxels, with a fixed shape and size across all datasets containing tumors. Vascular segmentation accuracy was quantified using the Dice similarity coefficient, computed between the post-processed vessel map and the corresponding ground-truth binary vessel mask. 
The mean and standard deviation of the Dice scores were reported across the dataset to characterize segmentation consistency. This dual evaluation framework enabled a nuanced understanding of the model’s ability to detect tumors and delineate vasculature. 

\begin{figure*}[h!]
\centering
\includegraphics[width=0.7\linewidth]{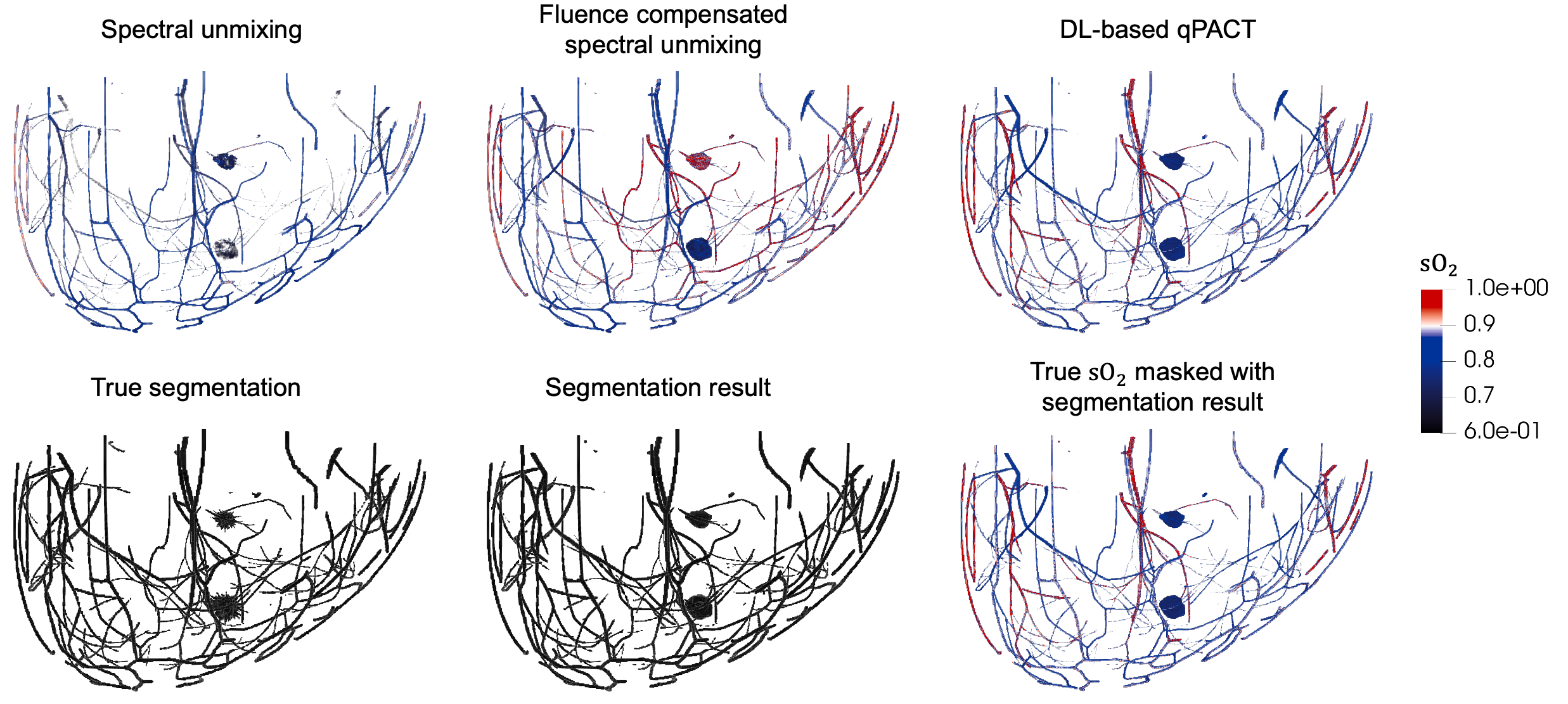}
\caption{Visual comparison of estimated blood oxygen saturation ($\text{sO}_2$) distributions and segmentation maps of vessels and tumors for the ID test set (skin color 1) in \textbf{Study 1}. Top row: estimated $\text{sO}_2$ maps obtained using spectral unmixing, fluence-compensated unmixing, and DL-based qPACT (left to right), each masked using the corresponding estimated segmentation map. Bottom row: true segmentation map (left), estimated segmentation map (center), and true $\text{sO}_2$ masked with the estimated segmentation map (right). DL-based qPACT provided more consistent $\text{sO}_2$ maps and more accurate segmentation.}
\label{visual_study_1}
\end{figure*}

\begin{figure*}[h!]
\centering
\includegraphics[width=0.8\linewidth]{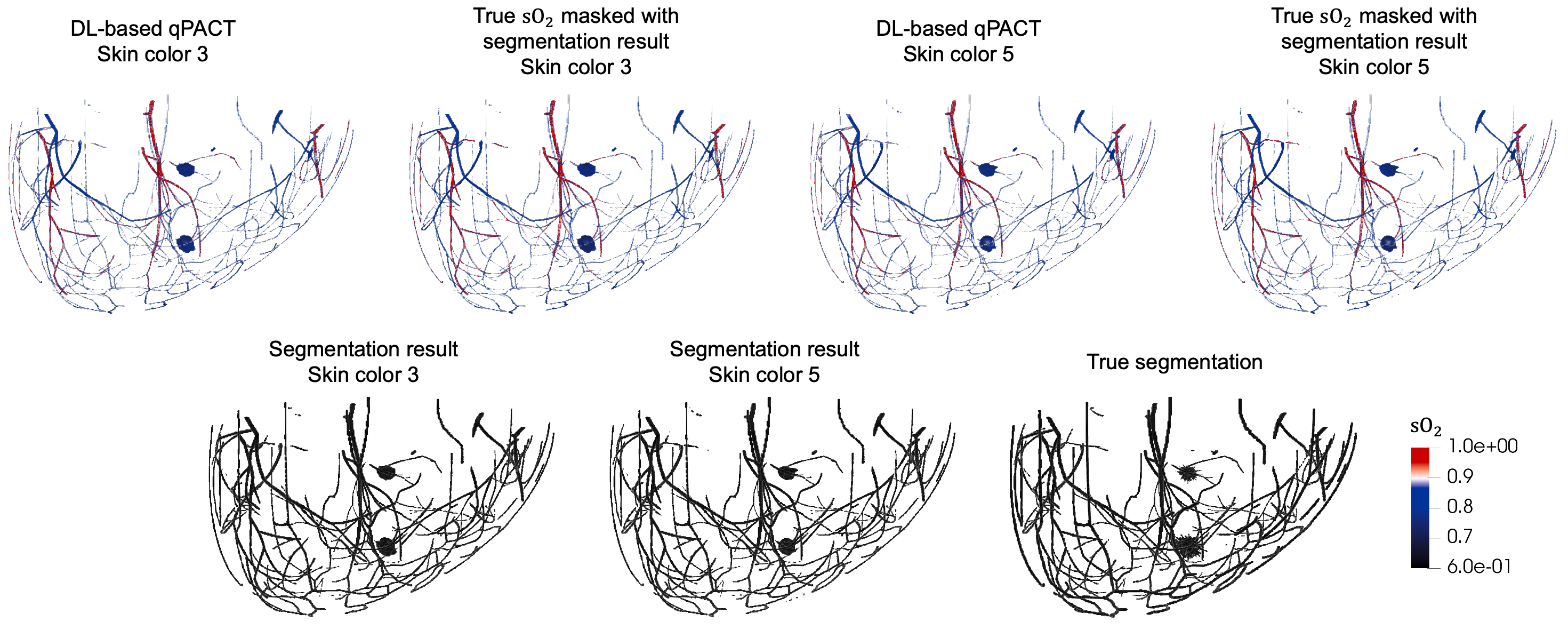}
\caption{Visual comparison of DL-based qPACT results for the OOD test sets (skin colors 3 and 5) 
in \textbf{Study 1}. Top row: estimated (first and third) and true (second and fourth) $\text{sO}_2$ maps, each masked with the corresponding estimated segmentation map, for skin color 3 (first and second) and skin color 5 (third and fourth). Bottom row: estimated segmentation masks for skin colors 3 (left) and 5 (center), and the corresponding true segmentation mask (right). DL-based qPACT maintained high visual fidelity in both segmentation and blood oxygenation estimates across diverse skin tones, demonstrating robust generalization.}
\label{visual_study_1_ood}
\end{figure*}

Quantitative assessment of the estimated $\text{sO}_2$ maps was performed separately for tumor and vascular regions, based on the network’s predicted segmentations. This approach reflects a clinically realistic scenario in which labeled segmentation maps are unavailable, and functional interpretation must rely directly on the model’s output. Similar evaluation strategies have been adopted in prior DL-based qPACT studies~\cite{bench2020toward,luke2019net,else2024effects}. Tumor $\text{sO}_2$ estimation was assessed by comparing the estimated average values within the model-identified tumor regions with the corresponding true average values. 
Vascular $\text{sO}_2$ estimation was evaluated as a function of depth to account for the exponential decay of optical fluence, which reduces the signal-to-noise ratio with increasing depth. Mean absolute error (MAE) was calculated at varying vessel depths to assess performance across the imaging volume.

For generalization assessment using the OOD test sets, the region within 0.6~mm depth from the skin surface was excluded from the outputs as post-processing. This region encompassed the epidermis, where melanin is concentrated. Because variations in melanin concentration determine skin tone, results in this superficial region can be comparatively inaccurate when the model encounters the test data with skin tones not represented in the training set. However, from a clinical perspective, the performance within the underlying breast tissue is of greater relevance than inaccuracies in the skin layer. Moreover, assuming skin thickness as prior knowledge is feasible. For these reasons, the superficial region was excluded when evaluating generalization with the OOD test sets.

\section{Results}
\label{results}

\subsection{Study-1 results}
Figure~\ref{visual_study_1} shows sample results for vessel and tumor segmentation, along with estimated $\mathrm{sO_2}$ distributions in blood vessels and tumors, 
under ID test conditions. 
Notably, the DL-based qPACT method produced $\mathrm{sO_2}$ estimates that more closely approximate the ground truth maps compared to the conventional approaches. 
The bottom row of Fig.~\ref{visual_study_1} shows the true and estimated segmentation masks, confirming that the DL-based qPACT method was able to localize vascular and tumor regions with high spatial fidelity. 
Figure~\ref{visual_study_1_ood} presents the corresponding results 
for OOD skin colors. 
The close alignment between the estimated and ground truth $\mathrm{sO_2}$ maps demonstrates the robust generalization of the DL-based qPACT method to the variations in skin tone. 

\begin{figure}[h!]
\centering
\includegraphics[width=0.99\linewidth]{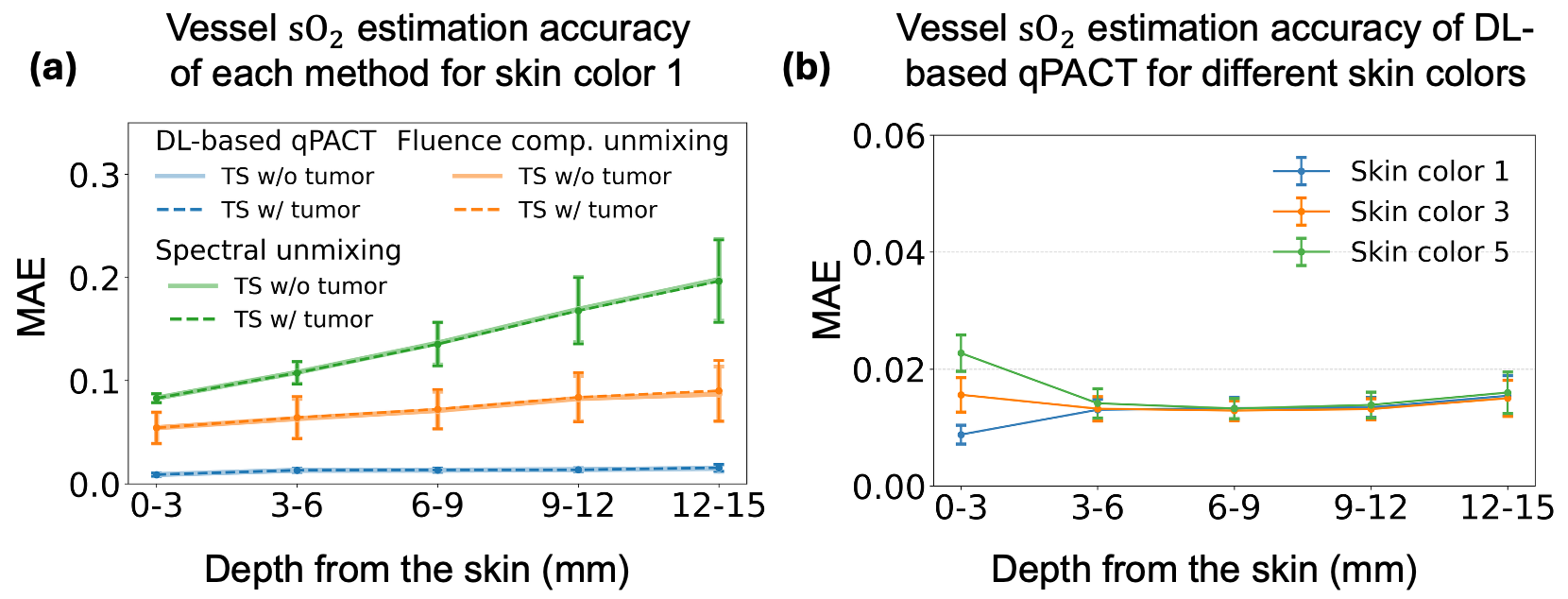}
\caption{Depth-wise MAE of estimated $\mathrm{sO_2}$ in segmented vessels for \textbf{Study 1}. (a) Comparison of spectral unmixing (green), fluence-compensated unmixing (orange), and DL-based qPACT (blue) on the ID test set (skin color 1). Results are shown separately for the test set with tumors (dashed, denoted TS w/ tumor) and the test set without tumors (solid, denoted TS w/o tumor). 
(b) Performance of DL-based qPACT across different skin colors: skin color 1 (blue), representing 
the ID case, and skin colors 3 (orange) and 5 (green), representing the OOD conditions. 
Error bars indicate standard deviation. 
DL-based qPACT maintained MAE below 3\% across all depths and skin tones, highlighting its robustness to depth-dependent fluence variations and distribution shifts.}
\label{study_1_vessel_so2}
\end{figure}

Figure~\ref{study_1_vessel_so2} shows the depth-wise accuracy of the estimated $\mathrm{sO_2}$ within vessels in Study 1. Panel (a) presents MAE values for ID skin color 1, evaluated on both tumor-absent and tumor-present test sets. The close agreement between these cases demonstrates that the presence of tumors does not substantially impact vascular $\mathrm{sO_2}$ estimation. Among the evaluated methods, the conventional spectral unmixing method exhibited considerable errors (close to 10\%) even at shallow depths (0 to 3~mm), with errors increasing at greater depths due to increased spectral coloring effect with optical attenuation. Fluence-compensated spectral unmixing showed improved performance, although it still showed 
noticeable accuracy reduction with depth. 
In contrast, the DL-based qPACT method consistently achieved lower errors (below 3\%) across all depths, demonstrating improved robustness against depth-dependent optical variations. Panel (b) further demonstrates that DL-based qPACT maintained comparably low errors across different skin tones, including OOD skin colors 3 and 5, suggesting robust generalization to OOD skin tone scenarios with minimal performance degradation.

Table~\ref{tab:study1_detection_results} presents tumor detection performance for Study 1. The DL-based qPACT method achieved high tumor detection accuracy for the ID test set (skin color 1), detecting 89 true positives with 6
false positives and 1 false negative. 
Additionally, the method maintained consistent performance on the OOD test sets, detecting 88 true positives in each case, with similarly low false positive and negative rates. These results suggest a reliable generalizability of the DL-based qPACT method.

\begin{table}[h]
\centering
\caption{Tumor detection results in \textbf{Study 1}.}
\label{tab:study1_detection_results}
\begin{adjustbox}{width=\columnwidth}
\begin{tabular}{lccc}
\toprule
\textbf{Test Set} & \textbf{True Positive} & \textbf{False Positive} & \textbf{False Negative} \\
\midrule
ID (skin color 1) & 89 & 6 & 1 \\
OOD-I (skin color 3) & 88 & 5 & 2 \\
OOD-II (skin color 5) & 88 & 3 & 2 \\
\bottomrule
\end{tabular}
\end{adjustbox}
\caption*{The total number of tumors present across all NBPs in each test set (number of true positives plus number of false negatives) is 90.}
\end{table}


\begin{figure}[h!]
\centering
\includegraphics[width=0.99\linewidth]{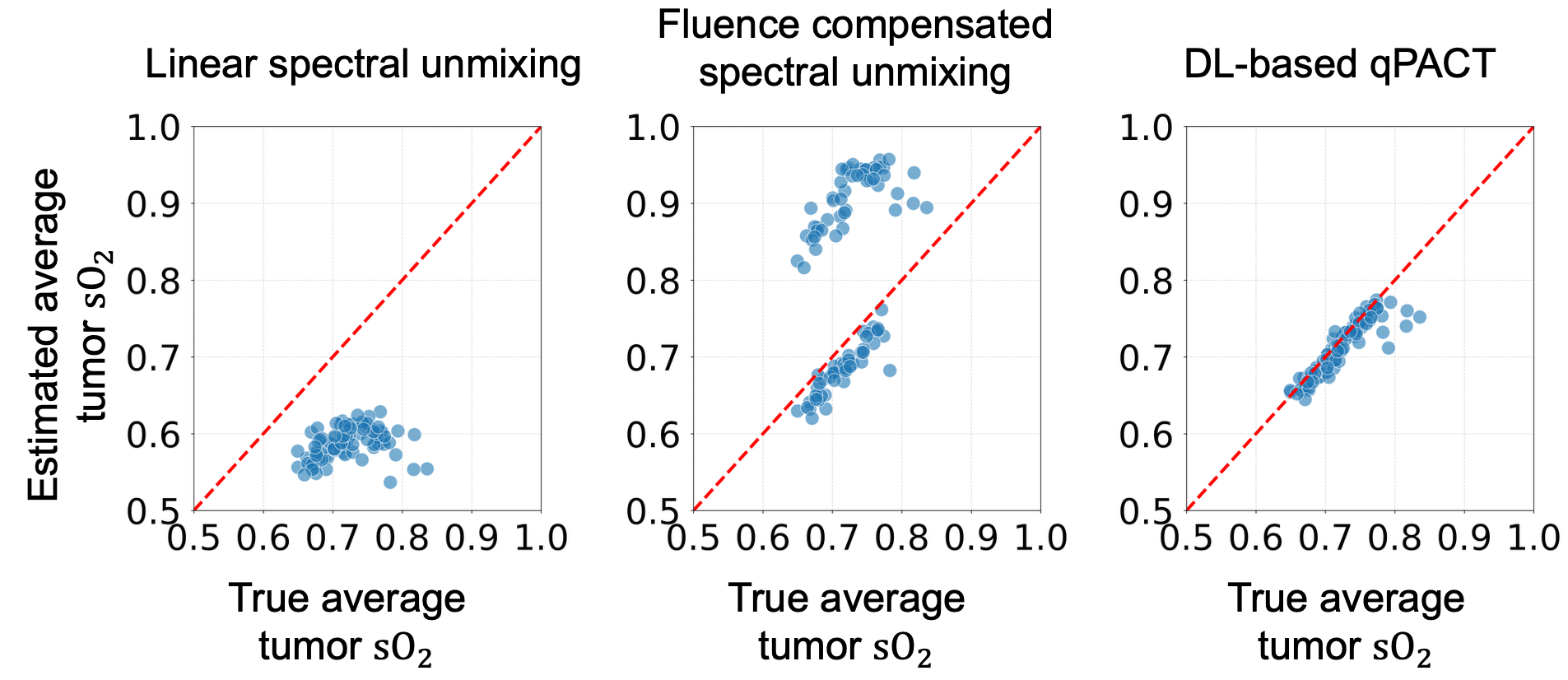}
\caption{Estimated vs. true average tumor $\mathrm{sO_2}$ values in \textbf{Study 1} for the ID test set (skin color 1). Scatter plots compare spectral unmixing (left), fluence-compensated unmixing (center), and DL-based qPACT (right). The red dashed line denotes the identity line, corresponding to 
perfect 
estimation. DL-based qPACT achieved the highest estimation accuracy, with estimates tightly clustering along the identity line, outperforming conventional methods.}
\label{average_tumor_case1_skin_color1}
\end{figure}

Figure~\ref{average_tumor_case1_skin_color1} provides a comparison of the methods in estimating average 
$\mathrm{sO_2}$ levels within tumors under the ID testing conditions. The scatter plots indicate that spectral unmixing consistently underestimated the true average values, whereas fluence-compensated unmixing showed reduced but still notable deviations from the true values. In contrast, DL-based qPACT estimates aligned closely with the true values, displaying minimal deviation and clustering tightly around the identity line. These results demonstrate 
the effectiveness of the DL-based qPACT method in quantitatively estimating tumor oxygenation under simplified acoustic conditions assumed in Study 1.

\begin{figure}[h!]
\centering
\includegraphics[width=0.7\linewidth]{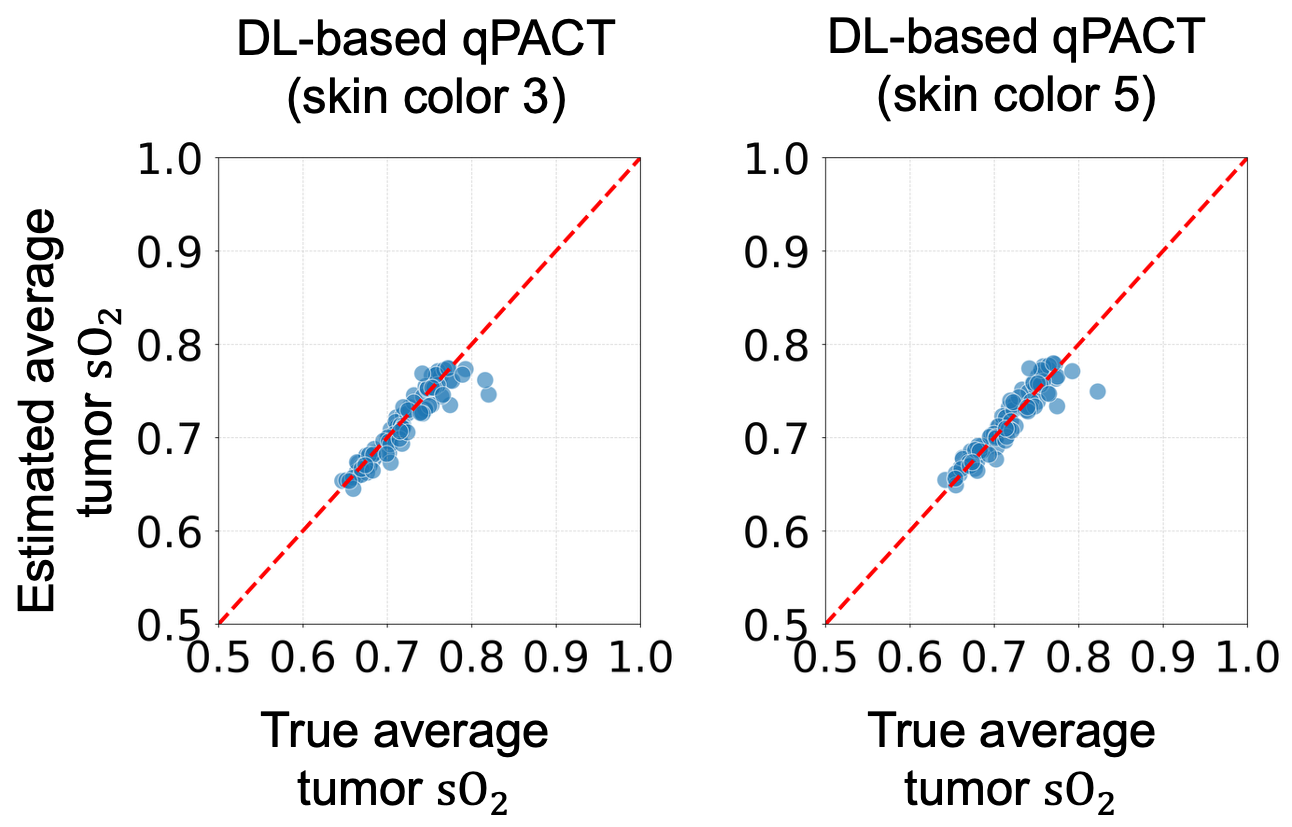}
\caption{Estimated vs. true average tumor $\mathrm{sO_2}$ values in \textbf{Study 1} for the OOD test sets (skin colors 3 and 5). 
Scatter plots show 
DL-based qPACT results for 
skin color 3 (left) and skin color 5 (right). 
DL-based qPACT demonstrated robust generalization, maintaining accurate tumor oxygenation estimates even under OOD conditions.} 
\label{average_tumor_case1_skin_color_ood}
\end{figure}

Figure~\ref{average_tumor_case1_skin_color_ood} presents 
the generalization performance of the DL-based qPACT method in estimating average 
$\mathrm{sO_2}$ within tumors for OOD skin tones in Study 1. Despite not being trained on skin colors 3 and 5, the method maintained a strong agreement between the estimated and true $\mathrm{sO_2}$ values, with points aligning closely along the identity line in both cases. This demonstrates high estimation accuracy with 
minimal bias introduced by variations in skin tone. 

\begin{figure*}[h!]
\centering
\includegraphics[width=0.7\linewidth]{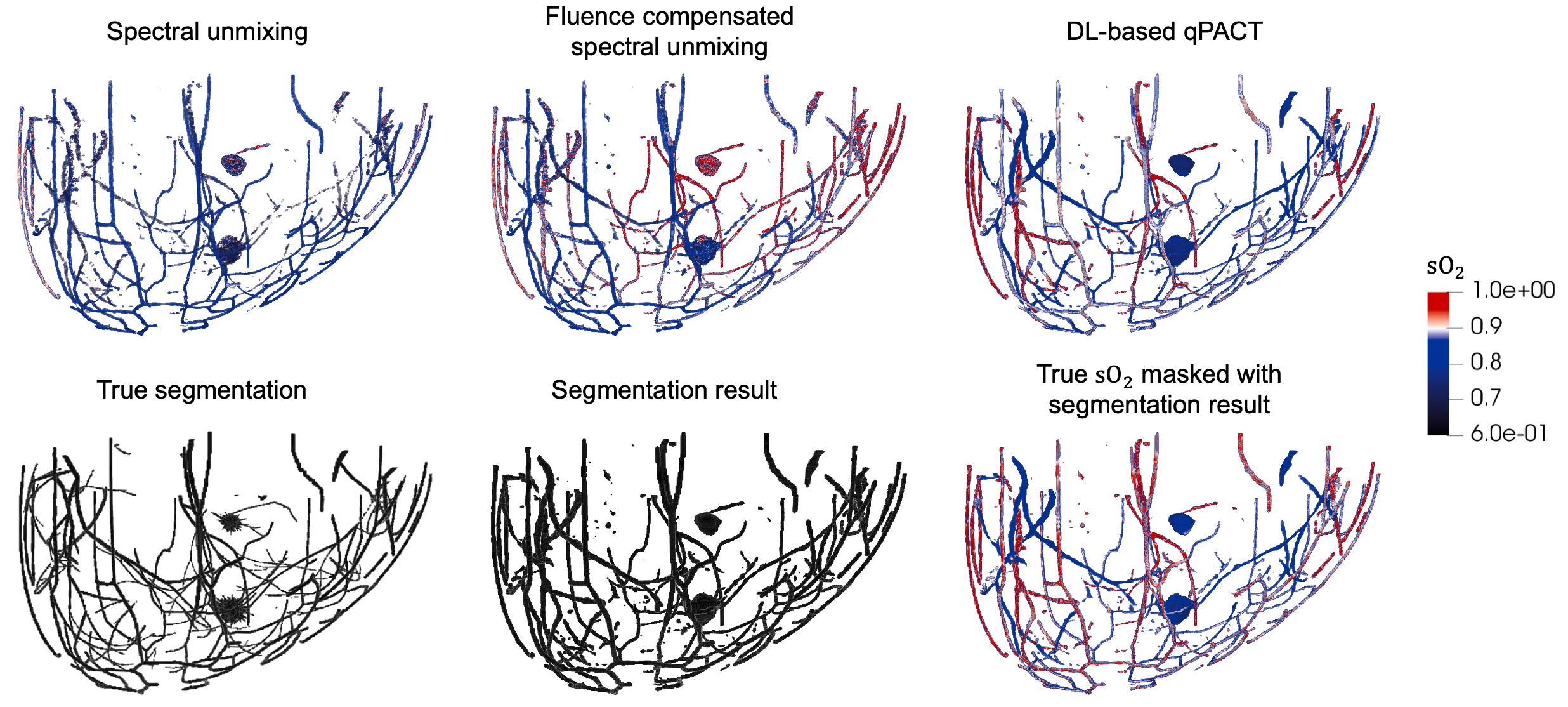}
\caption{Visual comparison of estimated $\text{sO}_2$ and segmentation maps of vessels and tumors for the ID test set (skin color 1) in \textbf{Study 2}. 
Top row: estimated $\text{sO}_2$ maps obtained using spectral unmixing, fluence-compensated unmixing, and DL-based qPACT (left to right). 
Bottom row: true segmentation map (left), estimated segmentation map (center), and true $\text{sO}_2$ masked with the estimated segmentation map (right). Despite reduced 
segmentation accuracy, DL-based qPACT preserved physiologically plausible $\text{sO}_2$ estimates, demonstrating robustness to errors in reconstructed initial pressure images caused by uncompensated acoustic heterogeneity.}
\label{visual_study_3}
\end{figure*}

\begin{figure*}[h!]
\centering
\includegraphics[width=0.8\linewidth]{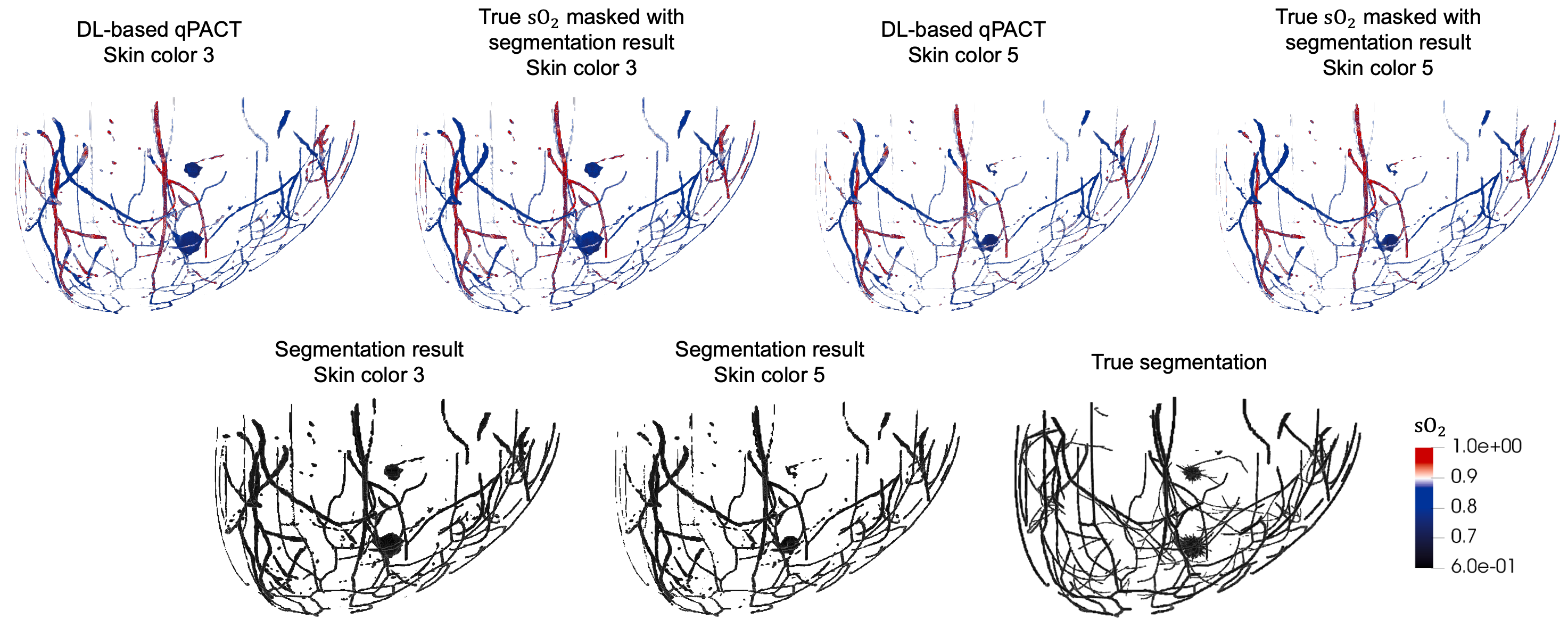}
\caption{Visual comparison of DL-based qPACT results for the OOD test sets (skin colors 3 and 5) in \textbf{Study 2}. Top row: estimated (first and third) and true (second and fourth) $\text{sO}_2$ maps, each masked with the corresponding estimated segmentation map, for skin color 3 (first and second) and skin color 5 (third and fourth). Bottom row: estimated segmentation masks for skin colors 3 (left) and 5 (center), and the corresponding true segmentation mask (right). DL-based qPACT maintained $\text{sO}_2$ estimation fidelity in detected regions, but showed declines in segmentation accuracy and sensitivity 
for OOD skin tones, underscoring potential challenges.}
\label{visual_study_3_ood}
\end{figure*}
The accuracy of 
vessel segmentations by the DL-based qPACT method, measured using the Dice coefficient, was highest for the ID test set (skin color 1), achieving 
0.8721$\pm$0.0094, which indicates 
strong overlap with the ground truth. Under OOD conditions, performance declined, with Dice scores of 0.7004$\pm$0.0266 for skin color 3 and 0.6985$\pm$0.0260 for skin color 5. Despite this reduction, 
the model maintained a reasonable level of performance, suggesting a certain 
degree of generalization to unseen skin tones.

\subsection{Study-2 results}
Figure~\ref{visual_study_3} shows an estimated segmentation mask 
and the corresponding estimated $\mathrm{sO_2}$ maps obtained with different methods, under ID conditions for Study 2. 
Despite the uncompensated acoustic heterogeneities in reconstructing 
the induced initial pressure, the DL-based qPACT method maintained its ability to produce accurate $\text{sO}_2$ estimates. 
The estimated segmentation maps exhibited greater structural fragmentation than in Study 1, 
likely due to artifacts resulting from modeling mismatches, yet the estimated $\text{sO}_2$ within the segmented regions remained consistent with the ground truth. These results indicate that the DL-based qPACT method retained its strength in $\text{sO}_2$ estimation, even though segmentation quality degrades under more realistic and challenging simulation conditions.

The visualizations in Figure~\ref{visual_study_3_ood} illustrate the challenges of generalization under the more realistic modeling conditions of Study 2. For skin color 3, the DL-based qPACT method continued to generate accurate $\text{sO}_2$ maps within detected tumor and vessel regions, although segmentation quality was visibly degraded. In the more challenging skin color 5 case, a tumor near the chest wall was entirely missed, likely due to reduced optical fluence and the resulting 
lower signal strength in this deeper region. For tumors that were successfully segmented, the estimated $\text{sO}_2$ values remained accurate, indicating the model's capacity to provide reliable oxygenation estimates. 

\begin{figure}[h!]
\centering
\includegraphics[width=0.99\linewidth]{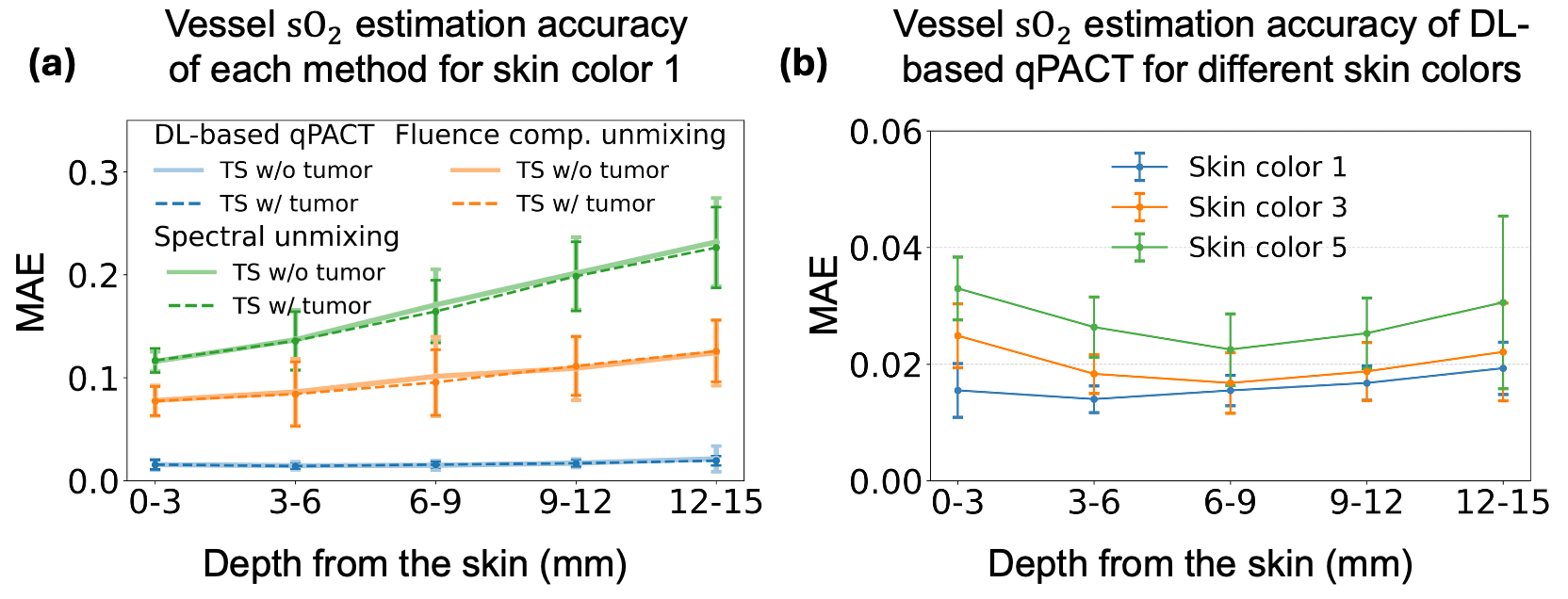}
\caption{Depth-wise MAE of estimated $\mathrm{sO_2}$ in segmented vessels for \textbf{Study 2}. (a) Comparison of spectral unmixing (green), fluence-compensated unmixing (orange), and DL-based qPACT (blue) 
on the ID test set (skin color 1). Results are shown separately for the test set with tumors (dashed, denoted TS w/ tumor) and the test set without tumors (solid, denoted TS w/o tumor). 
(b) Performance of DL-based qPACT across different skin tones: skin color 1 (blue), representing the ID case, and skin colors 3 (orange) and 5 (green), representing the OOD conditions. Error bars represent standard deviation. DL-based qPACT retained robust accuracy of vessel $\text{sO}_2$ estimates despite challenges posed by acoustic heterogeneity and distribution shifts.}
\label{study_3_vessel_so2}
\end{figure}

Figure~\ref{study_3_vessel_so2} shows 
the depth-wise MAE for the estimated $\mathrm{sO_2}$ within vessels in Study 2. Panel (a) presents MAE values under ID testing conditions, evaluated on both tumor-present and tumor-absent test sets. The close agreement between the results with these different test sets confirmed that tumor presence does not significantly influence $\mathrm{sO_2}$ estimation within the vessels for the considered methods.
Across all depths, the DL-based qPACT method outperformed both spectral unmixing and fluence-compensated unmixing.
Panel (b) displays the DL-based qPACT results for ID and OOD skin tones. 
Slightly higher estimation error observed in the shallow region (0–3~mm) relative 
to deeper regions (e.g., 3–6~mm) could possibly be attributed to 
acoustic heterogeneities at the interface between the acoustic coupling medium (water) and the breast tissue, which degrade signal quality near the surface. 
Nevertheless, panel (b) demonstrates that DL-based qPACT generalized well across skin tones in estimating vascular 
$\mathrm{sO_2}$, 
even under the more realistic simulation conditions of Study 2.

Table~\ref{tab:study3_detection_results} presents tumor detection performance for Study 2 across both ID and OOD skin tones. For the ID test set, the method achieved near-perfect results, 
with 89 true positives, only 2 false positives, and 1 false negative. However, detection performance declined under OOD testing conditions. For skin color 3, the number of true positives dropped to 77, accompanied by 13 false negatives. For skin color 5, the detection performance showed a more substantial decrease, with only 42 tumors detected and 48 missed. Although the false-positive rate remained relatively low across all skin tones, the decrease in true positive detection for the OOD skin tones indicates a reduction in sensitivity under increased distributional shift. This decline in sensitivity may stem from a combination of factors, including reduced optical fluence due to darker skin color and the absence of darker skin tones in the training data.

\begin{table}[ht]
\centering
\caption{Tumor detection results in \textbf{Study 2}.}
\label{tab:study3_detection_results}
\begin{adjustbox}{width=\columnwidth}
\begin{tabular}{lccc}
\toprule
\textbf{Test Set} & \textbf{True Positive} & \textbf{False Positive} & \textbf{False Negative} \\
\midrule
ID (skin color 1) & 89 & 2 & 1 \\
OOD-I (skin color 3) & 77 & 5 & 13 \\
OOD-II (skin color 5) & 42 & 6 & 48 \\
\bottomrule
\end{tabular}
\end{adjustbox}
\end{table}

\begin{figure}[h!]
\centering
\includegraphics[width=0.99\linewidth]{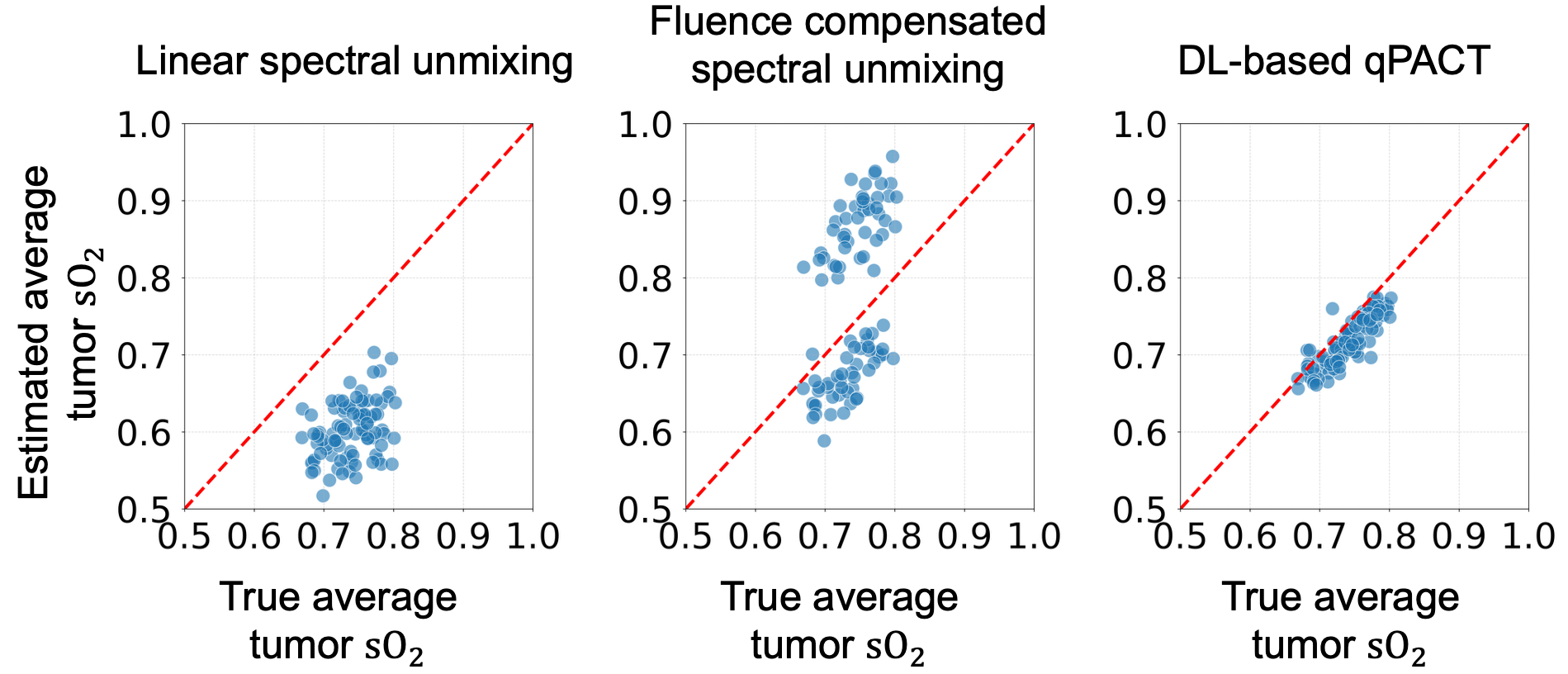}
\caption{Estimated vs. true average tumor $\mathrm{sO_2}$ values in \textbf{Study 2} for the ID test set (skin color 1). Scatter plots compare spectral unmixing (left), fluence-compensated unmixing (center), and 
DL-based qPACT (right). DL-based qPACT provided 
the most accurate estimates of average tumor $\mathrm{sO_2}$, indicating 
effective compensation for modeling errors in acoustic image reconstruction.}
\label{average_tumor_case3_skin_color1}
\end{figure}

\begin{figure}[h!]
\centering
\includegraphics[width=0.7\linewidth]{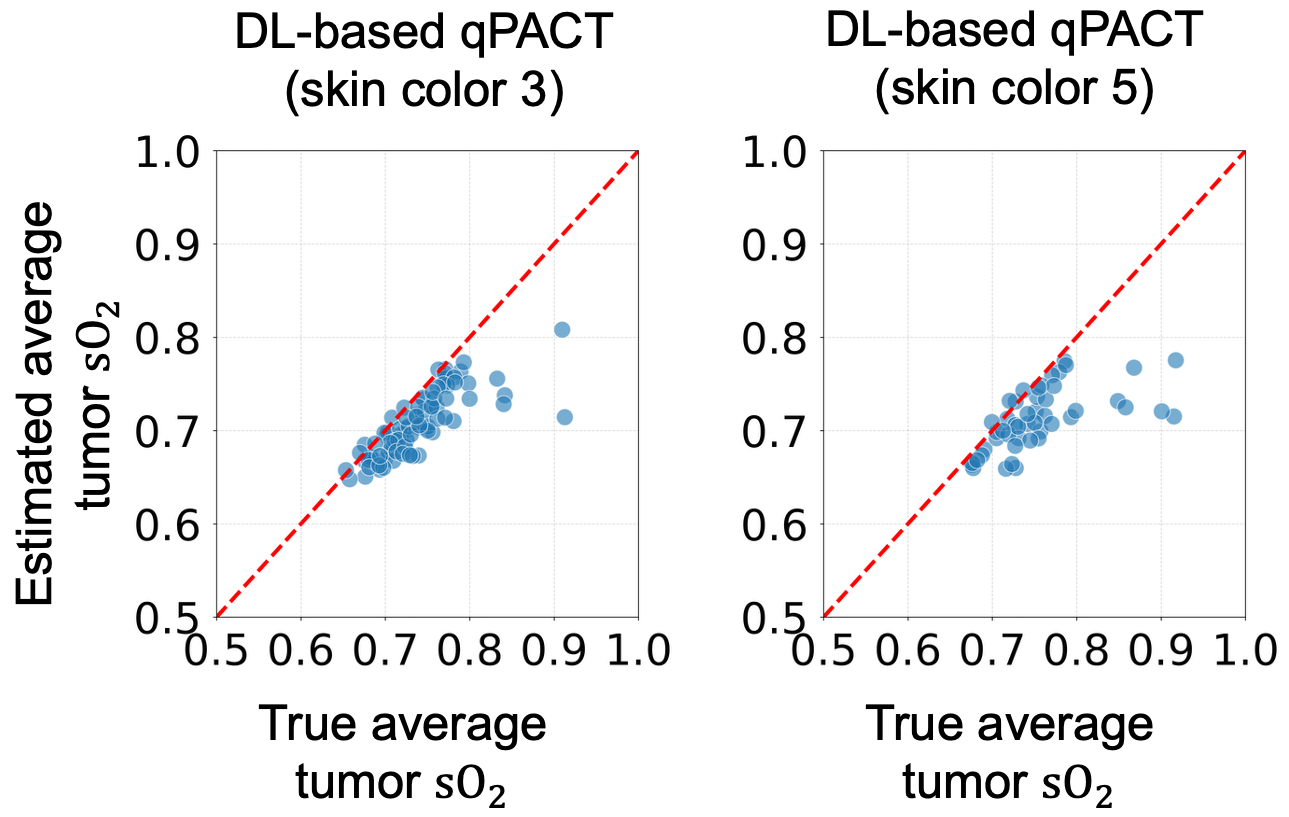}
\caption{Estimated vs. true average tumor $\mathrm{sO_2}$ values in \textbf{Study 2} for the OOD test sets (skin colors 3 and 5). 
Scatter plots show DL-based qPACT results 
for skin color 3 (left) and skin color 5 (right). Introducing more physiologically accurate acoustic properties in this study led to a notable decline in accuracy, particularly for skin color 5.} 
\label{average_tumor_case3_skin_color_ood}
\end{figure}

Figure \ref{average_tumor_case3_skin_color1} presents scatter plots comparing estimated and 
true average 
$\mathrm{sO_2}$ values in tumors under Study 2 for the ID test dataset. The conventional spectral unmixing method (left) significantly underestimated tumor $\mathrm{sO_2}$, exhibiting a clear downward bias and wide variability. Fluence-compensated spectral unmixing (center) improved the accuracy of estimated average tumor $\mathrm{sO_2}$ but still showed notable overestimates and dispersion relative to the identity line. In contrast, the DL-based qPACT method (right) demonstrates the closest agreement with the true values. 
This indicates that the DL-based qPACT method 
effectively mitigated modeling errors in the acoustic reconstruction, leading to more accurate tumor $\mathrm{sO_2}$ estimation.

Figure \ref{average_tumor_case3_skin_color_ood} illustrates the performance of the DL-based qPACT method in estimating the average tumor $\mathrm{sO_2}$ for OOD cases in Study 2. 
The method maintained reasonable accuracy for skin color 3, with the results moderately aligned around the identity line, whereas performance noticeably deteriorated for skin color 5. The scatter plot for skin color 5 revealed increased deviation from the identity line and greater variance, indicating a clear drop in tumor $\mathrm{sO_2}$ estimation accuracy. 

The accuracy of 
vessel segmentation 
markedly declined under the more realistic simulation conditions of Study 2. For the ID test set with skin color 1, the Dice coefficient dropped to 0.5268$\pm$0.0260, representing a substantial reduction compared to Study 1. Performance further deteriorated 
in OOD cases, with Dice scores of 0.4123$\pm$0.0257 for skin color 3 and 0.3864$\pm$0.0304 for skin color 5. These results suggest that the segmentation accuracy of 
the DL-based qPACT method diminished as the acoustic complexity increased, particularly for darker skin colors that were not represented in the training data.

\section{Discussion and Conclusion}
\label{discussion_and_conclusion}
This work demonstrates how realistic VI studies can be employed to systematically evaluate qPACT methods, revealing both their strengths and limitations under clinically relevant conditions. The employed framework leveraged 3D NBPs that incorporated anatomical, optical, and acoustic heterogeneity, enabling controlled yet physiologically realistic assessments. 
The VI framework was utilized to assess a representative DL-based qPACT method trained to jointly estimate $\mathrm{sO_2}$ and segment vascular and tumor regions from multispectral photoacoustic data. The evaluation spanned multiple sources of variability, including acoustic heterogeneity and 
distinct skin tones and demonstrated 
the impact of each on performance and generalization.

Results from the VI studies revealed that the considered DL-based qPACT method effectively estimated $\mathrm{sO_2}$ within tumors and vessels across different 
acoustic modeling assumptions in reconstructing 
the induced initial pressure. In ID test scenarios, the model maintained high accuracy in $\mathrm{sO_2}$ estimation, even as errors in initial induced pressure reconstructions increased from Study 1 to Study 2. Notably, Study 2 demonstrated that, despite reduced segmentation accuracy, 
the DL-based qPACT method was still able to estimate $\mathrm{sO_2}$ accurately under complex, clinically relevant acoustic and optical variability. This observation highlights the potential of DL-based qPACT frameworks to deliver accurate functional imaging in scenarios with complex clinically relevant variability.

However, the accuracy of the estimated $\mathrm{sO_2}$ and segmentation maps by the DL-based qPACT method for the OOD test sets with darker skin tones declined from Study 1 to Study 2.
While the method generalized well under the relatively simplified conditions of Study 1, its performance deteriorated under the more challenging conditions of Study 2. This was reflected in reduced tumor detection sensitivity, greater variability in $\mathrm{sO_2}$ estimates, and lower segmentation accuracy for darker skin tones not represented in the training data. These findings suggest that both physical factors, such as increased optical absorption and reduced signal-to-noise ratio in darker skin tones, and the lack of representative training data can limit model performance under clinically relevant distribution shifts. To ensure robust performance and applicability across a broad range of populations, it is essential to enhance 
training data diversity, particularly with respect to skin tone, and 
to account for the fundamental limitations imposed by the imaging physics. 

The observed discrepancy between robust ID performance and declining accuracy in OOD cases highlights a critical challenge in the development of DL-based qPACT methods. Evaluations conducted under 
oversimplified conditions 
can overestimate model performance, as they fail to incorporate 
the complexities of real-world anatomical and optical variations as well as inaccuracies in the estimated initial pressure distribution. The progressive decline in performance from Study 1 to Study 2 emphasizes the need for comprehensive validation pipelines that reflect clinical variability, including variations in skin color.


A persistent challenge in the field of qPACT is the lack of reliable \emph{in vivo} reference $\mathrm{sO_2}$ maps, which makes 
direct validation of reconstruction methods difficult. This limitation underscores the need for alternative evaluation strategies capable of yielding 
meaningful insights into method performance. 
The VI framework employed in this study provides such an alternative, enabling controlled and physiologically realistic assessments using realistic numerical phantoms.
While not a substitute for \emph{in vivo} validation, such VI studies are valuable tools for identifying method limitations, guiding algorithm development, and informing experimental design.

Overall, this study demonstrates the potential of the VI frameworks to evaluate the performance and robustness of qPACT methods in clinically relevant scenarios. By revealing both strengths and limitations of qPACT methods, VI studies can help ensure that future qPACT approaches are developed and validated with consideration for realistic anatomical and physiological variability.

\section*{Acknowledgements}
This work was supported in part by the National Institutes of Health, United States grants EB031585, EB034261 and EB031772. This work used the Delta system at the National Center for Supercomputing Applications through allocation MDE230007 from the Advanced Cyberinfrastructure Coordination Ecosystem: Services \& Support (ACCESS) program, which is supported by U.S. National Science Foundation grants \#2138259, \#2138286, \#2138307, \#2137603, and \#2138296.

\bibliographystyle{elsarticle-num}
\bibliography{references}

\appendix
\section{Loss Functions}\label{appendix-loss}
\subsection{Regression loss}\label{appendix-loss-regression}
The regression loss \(\mathcal{L}_{\text{reg}}\) is formulated as a weighted mean squared error (WMSE) between the estimated and true oxygen saturation distributions. Let \(\hat{y}_i\) represent the estimated $\text{sO}_2$ value at voxel \(i\), and \(y_i\) be the corresponding ground truth. To prioritize clinically relevant regions, a weight \(w_i^{\text{reg}} \geq 0\) is assigned to each voxel, with larger weights applied to 
voxels located within the outermost 1.5\,cm shell corresponding to vascular structures or viable tumor tissue. The WMSE loss is then defined as:
\begin{equation}
\mathcal{L}_{\text{reg}} \;=\; \frac{1}{N}\sum_{i=1}^{N} w^{\text{reg}}_i \,\bigl(\hat{y}_i - y_i\bigr)^2,
\end{equation}
where \(N\) is the total number of voxels in the output grid, corresponding to the discretized domain used for training and evaluation.



In this implementation, the image domain \(\Omega \subset \mathbb{R}^3\) corresponds to the spatial extent of the reconstruction volume, discretized into a uniform 3D voxel grid of size \(512 \times 512 \times 256\). Let \(\mathcal{I}\) denote the index set of all voxels 
in this grid. A voxel-wise weighting scheme is defined by partitioning \(\mathcal{I}\) into three disjoint subsets:
\[
\mathcal{I}_{\text{vas}} \;\subset\; \mathcal{I}, \quad
\mathcal{I}_{\text{vtc}} \;\subset\; \mathcal{I}, \quad
\mathcal{I}_{\text{bg}} \;=\; \mathcal{I} \,\setminus\, \bigl(\mathcal{I}_{\text{vas}} \cup \mathcal{I}_{\text{vtc}}\bigr).
\]
Here, \(\mathcal{I}_{\text{vas}}\) is the set of all voxels in the outermost 1.5\,cm shell 
corresponding to vascular structures, \(\mathcal{I}_{\text{vtc}}\) is the set of voxels in the same 
shell corresponding to 
viable tumor cells, and \(\mathcal{I}_{\text{bg}}\) represents the remaining voxels (i.e., background). Let
\[
N \;=\; \lvert \mathcal{I} \rvert, \quad
N_{\text{vas}} \;=\; \lvert \mathcal{I}_{\text{vas}} \rvert, \quad
N_{\text{vtc}} \;=\; \lvert \mathcal{I}_{\text{vtc}} \rvert, \quad
N_{\text{bg}} \;=\; \lvert \mathcal{I}_{\text{bg}} \rvert
\]
denote the cardinalities of these sets. The voxel-wise weight \(w^{\text{reg}}_i\) is then computed as
\begin{equation}
w^{\text{reg}}_i \;=\;
\begin{cases}
\dfrac{N}{N_{\text{bg}}}, & i \in \mathcal{I}_{\text{bg}}, \\[3ex]
\dfrac{N}{N_{\text{vas}}} \,\kappa, & i \in \mathcal{I}_{\text{vas}}, \\[3ex]
\dfrac{N}{N_{\text{vtc}}} \,\kappa, & i \in \mathcal{I}_{\text{vtc}},
\end{cases}
\label{eq:wmse}
\end{equation}
with \(\kappa \) as a scaling factor.
This weighting strategy amplifies the penalty for estimation errors in vascular and tumor-bearing regions confined to the outermost 1.5\,cm shell.


\subsection{Segmentation loss}\label{appendix-loss-segmentation}
The segmentation branch outputs a single-channel probability map \(\hat{s} \in [0,1]^{\mathcal{I}}\) over the discretized voxel grid. For each voxel index \(i \in \mathcal{I}\), \(\hat{s}_i\) denotes the estimated likelihood that the voxel belongs to a target structure, and \(s_i\) be the corresponding ground truth label. The segmentation loss \(\mathcal{L}_{\text{seg}}\) is defined as a weighted sum of the weighted binary cross-entropy (WBCE) loss and the soft Dice (sDICE) loss:
\begin{equation}
\mathcal{L}_{\text{seg}} \;=\; \mathcal{L}_{\text{WBCE}} \;+\; \beta\,\mathcal{L}_{\text{sDICE}},
\label{eqn:segmentation_loss}
\end{equation}
where \(\beta\) is a tunable hyperparameter that governs the relative importance of WBCE and sDICE, while the scalar $\eta$, defined in the main loss expression in Eq.~\eqref{eq:total_loss}, 
controls the overall contribution of the segmentation loss relative to the regression loss. Based on empirical evaluations in the numerical studies, the parameter values \(\eta = 0.03\) and \(\beta = 1.67\) were found to provide robust performance across test cases.



\paragraph{Weighted Binary Cross-Entropy (WBCE)}
Segmentation of small structures embedded within large background regions presents a well-known challenge in semantic segmentation, particularly when class imbalance and spatial context bias the network toward over-estimating the dominant class~\cite{milletari2016v, ronneberger2015u}. In this application, vessels (arteries and veins) and tumor structures (viable tumor cells and necrotic core) occupy relatively small volumes 
surrounded by counter-class voxels, making them prone to under-segmentation. To address 
this, a voxel-specific weight \(w^{\text{bce}}_i \geq 0\) was introduced into the binary cross-entropy formulation. In this implementation, the weighting scheme is determined by partitioning the domain \(\mathcal{I}\) into five disjoint sets:
\[
\begin{alignedat}{2}
    \mathcal{I}_{\text{art}}, \quad 
    \mathcal{I}_{\text{vein}}, \quad 
    \mathcal{I}_{\text{vtc}}, \quad 
    \mathcal{I}_{\text{nec}}, \quad \\[5pt]
    \mathcal{I}_{\text{else}} \;=\; 
    \mathcal{I} \setminus \left( 
        \mathcal{I}_{\text{art}} \cup 
        \mathcal{I}_{\text{vein}} \cup 
        \mathcal{I}_{\text{vtc}} \cup 
        \mathcal{I}_{\text{nec}} 
    \right),
\end{alignedat}
\]
where \(\mathcal{I}_{\text{art}}\) and \(\mathcal{I}_{\text{vein}}\) represent arterial and venous voxels, respectively, within the outermost 1.5\,cm shell, \(\mathcal{I}_{\text{vtc}}\) corresponds to 
viable tumor cells in the same shell, and \(\mathcal{I}_{\text{nec}}\) represents necrotic tissue within that shell. All remaining voxels are assigned to 
\(\mathcal{I}_{\text{else}}\). Let
\[
\begin{alignedat}{3}
    N              &= \bigl|\mathcal{I}\bigr|, \quad
    N_{\text{art}} &= \bigl|\mathcal{I}_{\text{art}}\bigr|, \quad
    N_{\text{vein}} &= \bigl|\mathcal{I}_{\text{vein}}\bigr|, \\[5pt]
    N_{\text{vtc}} &= \bigl|\mathcal{I}_{\text{vtc}}\bigr|, \quad
    N_{\text{nec}} &= \bigl|\mathcal{I}_{\text{nec}}\bigr|, \quad
    N_{\text{else}} &= \bigl|\mathcal{I}_{\text{else}}\bigr|.
\end{alignedat}
\]
denote the cardinalities of these sets.

The WBCE loss can be expressed as
\begin{equation}
\mathcal{L}_{\text{WBCE}} \;=\; -\frac{1}{N}\sum_{i \in \Omega} w^{\text{bce}}_i \,\Bigl[
s_i \ln(\hat{s}_i) \;+\; \bigl(1 - s_i\bigr)\,\ln\bigl(1 - \hat{s}_i\bigr)
\Bigr],
\end{equation}
where \(s_i \in \{0,1\}\) is the ground truth label for voxel \(i\), and \(\hat{s}_i\) is the corresponding estimated probability. The voxel-wise weight \(w^{\text{bce}}_i\) is assigned based on 
the tissue class:
\begin{equation}
w^{\text{bce}}_i \;=\;
\begin{cases}
\dfrac{N}{N_{\text{art}}}\gamma, & i \in \mathcal{I}_{\text{art}}, \\[2ex]
\dfrac{N}{N_{\text{vein}}}\gamma, & i \in \mathcal{I}_{\text{vein}}, \\[2ex]
\dfrac{N}{N_{\text{vtc}}}\gamma, & i \in \mathcal{I}_{\text{vtc}}, \\[2ex]
\dfrac{N}{N_{\text{nec}}}\gamma, & i \in \mathcal{I}_{\text{nec}}, \\[2ex]
\dfrac{N}{N_{\text{else}}}, & i \in \mathcal{I}_{\text{else}},
\end{cases}
\label{eq:seg_wbce}
\end{equation}
where \(\gamma\) is a parameter that modulates the weighting in the binary cross-entropy loss. This weighting scheme 
ensures that smaller, yet clinically significant regions are not overshadowed by adjacent, larger regions belonging to the opposing class.

\paragraph{Soft Dice (sDICE) Loss}
To further reinforce spatial overlap between the estimated segmentation \(\hat{\mathbf{s}}\) and the ground truth \(\mathbf{s}\), a differentiable variant of the Dice similarity coefficient, referred to as 
sDICE, 
is employed~\cite{milletari2016v}:
\begin{equation}
\text{sDICE}(\hat{\mathbf{s}}, \mathbf{s}) \;=\; \frac{2\,\sum_{i\in\mathcal{I}} \hat{s}_i s_i}{\sum_{i\in\mathcal{I}} \hat{s}_i + \sum_{i\in\mathcal{I}} s_i},
\end{equation}
with the corresponding sDICE 
loss defined as: 
\begin{equation}
\mathcal{L}_{\text{sDICE}} \;=\; 1 - \text{sDICE}(\hat{\mathbf{s}}, \mathbf{s}).
\end{equation}
This additional loss term complements the WBCE loss by placing greater emphasis on the overall overlap of target structures, thereby encouraging 
accurate boundary delineation and spatial coherence. 

\subsection{Loss function curriculum}\label{appendix-loss-curriculum}
During the training process, the weight factor \(\kappa\) in \(w^{\text{reg}}_i\) for the regression loss was empirically set to 10 based on our experiments. The parameter \(\gamma\) in \(w^{\text{bce}}_i\) 
was scheduled with values \(\{1, 0.5, 0.25\}\), where each \(\gamma\) value corresponded to a training phase consisting of 200 epochs, resulting in a total of 600 training epochs.. This progressive adjustment initially emphasized clinically targeted regions and subsequently reduced their relative importance, enabling gradual refinement of 
the network’s segmentation performance. 
Such a strategy has been found 
to enhance delineation between clinically significant regions and the background by allowing the model to adaptively focus on feature refinement over the course of training. 


\end{document}